\documentclass{article}
\usepackage{graphicx}

\begin{document}

\def\pa{\partial}
\def\Bb{{\bf B}}
\def\jb{{\bf j}}
\def\Ab{{\bf A}}
\def\Bp{B_{\phi}}
\def\Rs{R_{\odot}}
\def\ol{\overline}

\title{ON THE CONNECTION BETWEEN MEAN FIELD DYNAMO THEORY AND FLUX TUBES}
\author{Arnab Rai Choudhuri (arnab@physics.iisc.ernet.in)\\
Department of Physics, Indian Institute of Science}
\date{}
\maketitle 
\begin{abstract}

Mean field dynamo theory deals with various mean quantities and does not directly
throw any light on the question of existence of flux tubes.  We can, however, draw
important conclusions about flux tubes in the interior of the Sun 
by combining additional arguments with the insights gained from solar dynamo
solutions. The polar magnetic field of the Sun is of order 10 G, whereas the 
toroidal magnetic field at the bottom of the convection zone has been estimated
to be 100,000 G.  Simple order-of-magnitude estimates show that the shear in
the tachocline is not sufficient to stretch a 10 G mean radial field 
into a 100,000 G mean
toroidal field.  We argue that the polar field of the Sun must get concentrated
into intermittent flux tubes before it is advected to the tachocline.  We estimate
the strengths and filling factors of these flux tubes.  Stretching by shear in
the tachocline is then expected to produce a highly intermittent magnetic
configuration at the bottom of the convection zone.  The meridional flow at
the bottom of the convection zone should be able to carry this intermittent
magnetic field equatorward, as suggested recently by Nandy and Choudhuri (2002).
When a flux tube from the bottom of the convection zone rises to a region of pre-existing
poloidal field at the surface, we point out that it picks up a twist in accordance
with the observations of current helicities at the solar surface.   

\end{abstract}
\section{Introduction}

Observations of the solar surface clearly indicate that the magnetic field
there exists in the form of flux tubes.  We see magnetic flux concentrations
of various sizes, from large sunspots to fibril flux tubes at the limit
of seeing. There is no direct observational evidence whether the magnetic field
exists in the form of flux tubes even in the interior of the convection
zone.  Various theoretical considerations, however, suggest that this must
be so due to the interaction of the magnetic field with the surrounding
convection.  Simulations of flux tube rise explain various aspects of bipolar
active regions on the surface rather well, suggesting that the magnetic field
probably rises as flux tubes from the bottom of the convection zone (Choudhuri,
1989; D'Silva and Choudhuri, 1993; Fan, Fisher, and DeLuca, 1993; Caligari,
Moreno-Insertis, and Sch\"ussler, 1995; Longcope and Fisher, 1996; Longcope 
and Choudhuri, 2002).

One of the important problems in solar physics is to understand the generation
of the solar magnetic field by the dynamo process (for
a recent review, see Choudhuri, 2002).  Since doing a full dynamical
simulation of the dynamo process in the entire solar convection zone is an
extremely difficult problem (Gilman, 1983; Glatzmaier, 1985), most of the solar
dynamo models are of kinematic nature and are based on the mean field dynamo
equation (Moffatt, 1978, Ch.\ 7; Choudhuri, 1998, Ch.\ 16).  This mean field 
equation is obtained by averaging over the fluctuating magnetic and velocity
fields.  If the magnetic field exists in the form of flux tubes, then certainly
the magnetic fluctuations are much larger than the mean value.  One can raise
doubts whether the averaging procedure and the subsequent approximations like
the {\em first order smoothing approximation} can be trusted in such a situation.
The full dynamical simulations have demonstrated that the dynamo process
really does take place (Gilman, 1983; Glatzmaier, 1985), even though these simulations
failed to yield realistic models of the solar dynamo.  We, therefore, believe
that the mean field equation captures the essence of the dynamo process in
some approximate way, even though it may be difficult to justify it 
rigorously on mathematical grounds. We shall, however, show that if one blindly
follows the results of mean field theory without keeping in mind that the 
magnetic field exists in the form of flux tubes, then one is often drawn
into misleading conclusions.

As of now, there exists no mean field formulation which addresses the issue
of flux tubes. How should we then reconcile the results of mean field dynamo
theory with the existence of flux tubes?  The aim of this paper is to suggest
the following two-step procedure:
\begin{enumerate}
  \item First solve the mean field dynamo equation to get a qualitative idea of
  how the mean magnetic field behaves;
  \item Then use other basic physics considerations to figure out how the magnetic
  field may be structured in flux tubes in different regions, leading to a more
  complete picture of the dynamo process.
\end{enumerate}
The first step of solving the mean field equation will not be presented in this
paper.  We shall rather rely on the insight gained from the dynamo calculations
presented in several recent papers (Choudhuri, Sch\"ussler, and Dikpati,
1995; Durney, 1995, 1996, 1997;
Dikpati and Charbonneau, 1999; K\"uker, R\"udiger, and Schultz, 2001; Nandy 
and Choudhuri, 2001, 2002).
We shall explicitly carry out the second step listed above 
in this paper, based on the general 
picture of the mean magnetic field that emerges from the above dynamo calculations.
In spite of some differences in the approaches of the above authors, there are
many common characteristics.  Since the meridional circulation plays a crucial role
in the works of all the above authors, we shall refer to this type of dynamo
model as the circulation-dominated solar dynamo model or CDSD model in brief.

Before carrying the second step listed above, let us summarize the general 
characteristics of the CDSD model on which we
shall carry out our procedure.  The toroidal field is produced by the stretching
of poloidal field lines by differential rotation.  Since helioseismology has shown
that differential rotation is concentrated in the tachocline at the base of the
convection zone, there is virtually universal agreement that the strong toroidal
field must be produced in this tachocline. This toroidal field must rise from there
to produce active regions on the solar surface.  Simulations of this rise suggest
that the strength of this toroidal field at the base of the convection zone must
be of order 100,000 G (Choudhuri and Gilman, 1987; Choudhuri,
1989; D'Silva and Choudhuri, 1993; Fan, Fisher, and DeLuca, 1993).  
Since convective turbulence cannot twist such strong fields, the conventional
$\alpha$-effect (Parker 1955; Steenbeck, Krause, and R\"adler 
1966) is unlikely to be operative.
All authors working on the CDSD model have invoked the idea of Babcock (1961) and
Leighton (1969) that the poloidal field is produced near the solar surface by
the decay of tilted active regions.  The poleward meridional circulation in the
upper regions of the convection zone has been mapped to a depth of about 15\% of
the solar radius (Giles, Duvall, and Scherrer, 
1997; Braun and Fan, 1998).  To conserve mass,
there has to be an equatorward return flow through the lower layers of the 
convection zone.  The poloidal field generated at the surface is carried by this
meridional circulation first poleward and then down underneath to the tachocline,
where it can be stretched to produce the toroidal field.  All authors working on
the CDSD model agree on this general scenario and many of our deductions in this
paper are based on this generally agreed scenario.

Nandy and Choudhuri (2002) have recently introduced a new idea in the CDSD model
and later in this paper we shall explore some of the consequences of this idea.
So we summarize this new idea here. The differential rotation $d \Omega/ dr$, which
has a negative value in the high latitudes within the tachocline, is much stronger
there compared to what it is in the low latitudes within the tachocline, where
it is positive (see Fig.\ 1 of Nandy and Choudhuri, 2002).  A differential rotation
of this kind tends to produce the strong toroidal field (and thereby enhanced
magnetic activity) at high latitudes rather than at low latitudes where sunspots
are seen (Durney, 1997;  Dikpati and Charbonneau, 1999; K\"uker, R\"udiger, and
Schultz, 2001; Nandy and 
Choudhuri, 2002).  Since the meridional circulation is generally believed to be 
driven by the turbulent stresses of the convection zone, most authors working on
the CDSD model assume the meridional circulation to be confined within the convection
zone.  In view of the fact that nothing much is known about the equatorward return
flow of this circulation in the deeper layers of the convection zone---either from
observations or from any well-established theoretical model, Nandy and Choudhuri
(2002) have proposed the tentative hypothesis that this flow penetrates slightly
below the convection zone, to a greater depth
than usually believed.  Even if the strong toroidal field is produced at high
latitudes within the tachocline, this flow would take it to the stable layers
below the convection zone and would not allow it to emerge at high latitudes.
The toroidal field is then transported equatorward by the meridional circulation
through the stable layers and comes within the convection zone only at low
latitudes where the meridional flow rises.  Thus, even though the toroidal
field is produced in the tachocline at high latitudes, it comes within the convection
zone and becomes buoyant only at low latitudes, thereby ensuring that flux eruption
takes place only at the typical sunspot latitudes.  We shall show that this idea
has some additional attractive features when we introduce flux tube considerations
with it.

We show in the next section that the shear in the tachocline is insufficient to
stretch a 10 G mean polar field into a 100,000 G mean toroidal field.  This fact
forces us to conclude that the polar field must become intermittent before being
advected to the tachocline.  The geometry of the polar field before entering
the tachocline is worked out in \S3. Then \S4 looks at the implications of an
intermittent field specifically for the model of Nandy and Choudhuri (2002).
Some properties of flux tubes at the bottom of the convection zone are discussed
in \S5. The question of magnetic helicity and twists of flux tubes is addressed
in \S6. Finally, some of the main points are summarized in the concluding section.

\section{The stretching of field lines by differential rotation}

Magnetogram observations show that the maximum radial field $B_r$ in the polar
region can be of order 10 G.  According to the CDSD model, this field is transported
below by the meridional circulation sinking near the poles and then brought to
the tachocline to be stretched by the differential rotation.  Although the field
is taken to regions of much higher density and any horizontal component will be
compressed by a large factor in this process, a radial field remains more or 
less unaffected when it is transported by a nearly vertically downward flow.  We
expect that the maximum value of $B_r$ even at the bottom of the convection zone
in the polar region to be about 10 G.  This field is stretched by the differential
rotation in the tachocline to generate the toroidal component.  The full equation
describing the evolution of the toroidal component is 
\begin{eqnarray}
\lefteqn{\frac{\pa \Bp}{\pa t} 
+ \frac{1}{r} \left[ \frac{\pa}{\pa r}
(r v_r \Bp) + \frac{\pa}{\pa \theta}(v_{\theta} \Bp) \right]
= \eta \left( \nabla^2 - \frac{1}{s^2} \right) \Bp } \nonumber \\ 
 & &  + s(\Bb_p.\nabla)\Omega - \nabla\eta \times \left(\nabla \times 
\Bp \bf{e_\phi}  \right),
\end{eqnarray}
where $s = r \sin \theta$ and the other symbols have the usual meanings.  It is
the term $s(\Bb_p.\nabla)\Omega$ which describes the generation of the toroidal
field from the poloidal field due to the stretching by differential rotation.
Within the tachocline, the time evolution of the toroidal field is dominated by
this term.  So the toroidal field generated in the tachocline is approximately
given by 
$$ \Bp \approx s(\Bb_p.\nabla)\Omega \, \tau, \eqno(2) $$
where $\tau$ is the time for which the stretching takes place.  Since the differential
rotation is mainly radial in the tachocline, (2) leads to
$$\frac{\Bp}{B_r} \approx s \frac{\Delta \Omega}{\Delta r} \tau. \eqno(3) $$
Let us now estimate the approximate value of the ratio $\Bp/B_r$ of the toroidal
field to the poloidal field from which it is produced.
Since the toroidal field $\Bp$ is believed to have the rather high value of 100,000 G,
we use such approximate values of various quantities in (3) which would make
the ratio $\Bp/B_r$ as large as possible.  Within the tachocline, $\Omega$ changes
from about 2150 nHz at the top to 
about 2800 nHz at the bottom, so we take $\Delta \Omega \approx 650$
nHz.  Let us use the rather large value $s \approx 0.5 \Rs$ and the rather small value
$\Delta r \approx 0.1 \Rs$ for the thickness of the tachocline. On taking $\tau \approx
10$ yr and substituting all these values in (3), we get
$$\frac{\Bp}{B_r} \approx 1000. \eqno(4)$$
Thus the radial magnetic field of 10 G can be stretched to produce a toroidal field
of maximum strength 10,000 G---smaller by a factor of 10 compared to
the strength 100,000 G inferred from the flux tube simulations.  To produce a toroidal
field of 100,000 G by the stretching of the radial field, we need to begin with a
radial field of order 100 G.

It is clear that a radial field of about 10 G cannot be stretched in the tachocline
to produce a toroidal field of 100,000 G, as required to match surface observations
in flux tube simulations.  What could be amiss here?  We now have to be careful not
to mix up mean field arguments with flux tube arguments.  The value 10 G for the radial
field is the mean value appropriate for mean field theories.  Although flux tube
simulations tell us that the magnetic field inside a flux tube has to be 100,000 G,
these simulations do not throw any light on the volume filling factor of these
flux tubes at the bottom of the convection zone or on the mean value of the toroidal
field there. Our equations (1)--(4) all hold for mean values.  If a mean radial field
of 10 G produces a mean toroidal field of 10,000 G, which is intermittently
concentrated into flux tubes with field 100,000 G, then all our theoretical requirements
are satisfied.  This gives us a volume filling factor of about $f \approx 0.1$
for the toroidal field.  Keeping in mind
that our estimate of $\Bp/B_r$ is based on values of various things which make this
ratio maximum, we conclude that 0.1 is the upper limit of the filling factor $f$.  The
actual filling factor may be somewhat smaller than this, but probably not smaller by
an order of magnitude.  We thus see that we can derive an approximate value of the
volume filling factor of the toroidal field from very simple considerations.  An estimate 
of the volume filling factor would be of interest for various purposes.  Within the
next few years, it may be possible to use helioseismology to probe the magnetic field
at the bottom of the convection zone.  Apart from the theoretical expectation that the
magnetic field inside flux tubes should be of order 100,000 G, an idea of the volume
filling factor would give helioseismologists a good clue of what to look for. 

There are two possible routes through which we may obtain toroidal flux tubes with
100,000 G field starting from a radial field of 10 G:
\begin{enumerate}
\item A fairly uniform $B_r \approx 10$ G may first give rise to a fairly uniform
$\Bp \approx 10,000$ G, which then gets broken into toroidal flux tubes with
magnetic field 100,000 G.
\item A fairly uniform $B_r \approx 10$ G may first get broken into vertical flux
tubes with field of order 100 G and then these vertical flux tubes may be stretched
in the tachocline to produce toroidal flux tubes with field 100,000 G.
\end{enumerate}
We here argue in favour of the second possibility. The magnetic field 100,000 G inside
the toroidal flux tubes is much larger than the equipartition magnetic field (i.e.\ 
the magnetic field having energy density equal to the kinetic energy density of
convective turbulence) at the tachocline.  Convective turbulence certainly could
not concentrate the magnetic field into flux tubes of such intensity.  On the other
hand, convective turbulence above the tachocline could easily concentrate the radial
field into vertical flux tubes with field 100 G (considerably less than the
equipartition value) and then these flux tubes could
be stretched by the differential rotation to produce toroidal flux tubes with field
100,000 G. It is not difficult to suggest a physical scenario how the formation of
the vertical flux tubes above the tachocline may come about.  In a region of 
convection, magnetic fields tend to get concentrated in the boundaries of convection
cells.  This is seen at the solar surface as well as in simulations of magnetoconvection.
We believe that the convection cells deeper down in the solar convection zone are
bigger in size, since the scale height is larger there.  So a nearly uniform
radial field at the surface is pushed by the downward meridional flow into regions
where convection cells are larger and consequently cell boundaries are further and
further apart.  The magnetic field would be pushed to these cell boundaries and would
become highly intermittent.  A snapshot of magnetic field lines in the vertical plane
may appear as shown in Figure~1.

\begin{figure}
\centerline{\includegraphics[height=7cm,width=7cm]{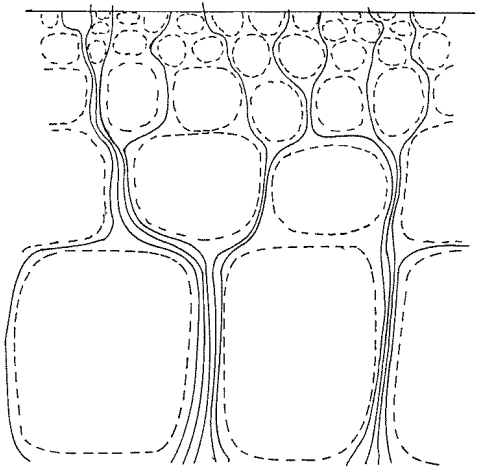}}
\caption{A sketch of magnetic field lines in a vertical plane near 
the polar region of the solar
convection zone.  The dashed lines indicate convection cells, which become
larger as we go deeper.}
\end{figure}

\section{Probable magnetic field geometry above and within the tachocline}

We now try to figure out what the magnetic field may look like above and within
the tachocline. Earlier, we had put such values of various quantities in (3) that
would make the ratio $\Bp/B_r$ as large as possible.  A more reasonable value of
this ratio may be
$$\frac{\Bp}{B_r} \approx 200 \eqno(5)$$
rather than 1000 as given in (4). In that case, we have to begin 
with vertical flux tubes having inside fields $B_r \approx 500$ G  which can be
stretched in the tachocline to produce toroidal flux tubes with magnetic fields
of 100,000 G inside. Since the mean value of $B_r$ is 10 G, we need to have a
filling factor of only $f = 0.02$ or $2\%$ if this mean field has to 
be concentrated into flux tubes of strength 500 G.

We can also draw some conclusions about the sizes of these flux tubes.  We are
arguing that these radial flux tubes above the tachocline in the polar region
get stretched in the tachocline to produce the toroidal flux tubes and then
parts of these flux tubes rise to produce the active region.  Therefore, the
magnetic flux associated with a vertical flux tube above the tachocline would be 
the flux which we would finally find as the flux in a typical sunspot, i.e.
$$ B_{ft} \, r_{ft}^2 = B_{ss} \,  r_{ss}^2. $$
Here $B_{ft}$ is the magnetic field inside a vertical flux tube above the
tachocline (estimated above to be 500 G) and $B_{ss}$ is the magnetic field
inside a sunspot at the photospheric level (3000 G), whereas $r_{ft}$ and
$r_{ss}$ are the corresponding radii.  We then have 
$$r_{ft} = r_{ss} \left( \frac{B_{ss}}{B_{ft}} \right)^{1/2}. \eqno(6)$$
The typical radius of a flux tube above
the tachocline thus has to be somewhat larger than 
the radius of a sunspot at the photospheric
level.  On taking the radius of a typical sunspot to be about 5000 km,
(6) gives 
$$r_{ft} \approx 12,000 \; \mbox{km}. \eqno(7)$$
In order to have a filling factor of about $f$,
these flux tubes need to have typical horizontal separations of order
$$s \approx f^{-1/2} \, r_{ft}, \eqno(8)$$
which turns out to be
$$s \approx 85,000 \; \mbox{km} \eqno(9)$$ 
on using
$f \approx 0.02$.  This is of the same order as the vertical scale height at the
bottom of the solar convection zone and one would naively expect the convection
cells to have horizontal sizes of this order in that region.  Our arguments thus
lead us to the conclusion that the typical horizontal separations of these
flux tubes must be of the same order as the horizontal sizes of the convection
cells---a very sensible conclusion in view of the fact that we expect the flux
tubes to be concentrated at the boundaries of convection cells.  This remarkable
conclusion gives us the confidence that we are approximately on the correct
track.

\begin{figure}
\centerline{\includegraphics[height=7cm,width=7cm]{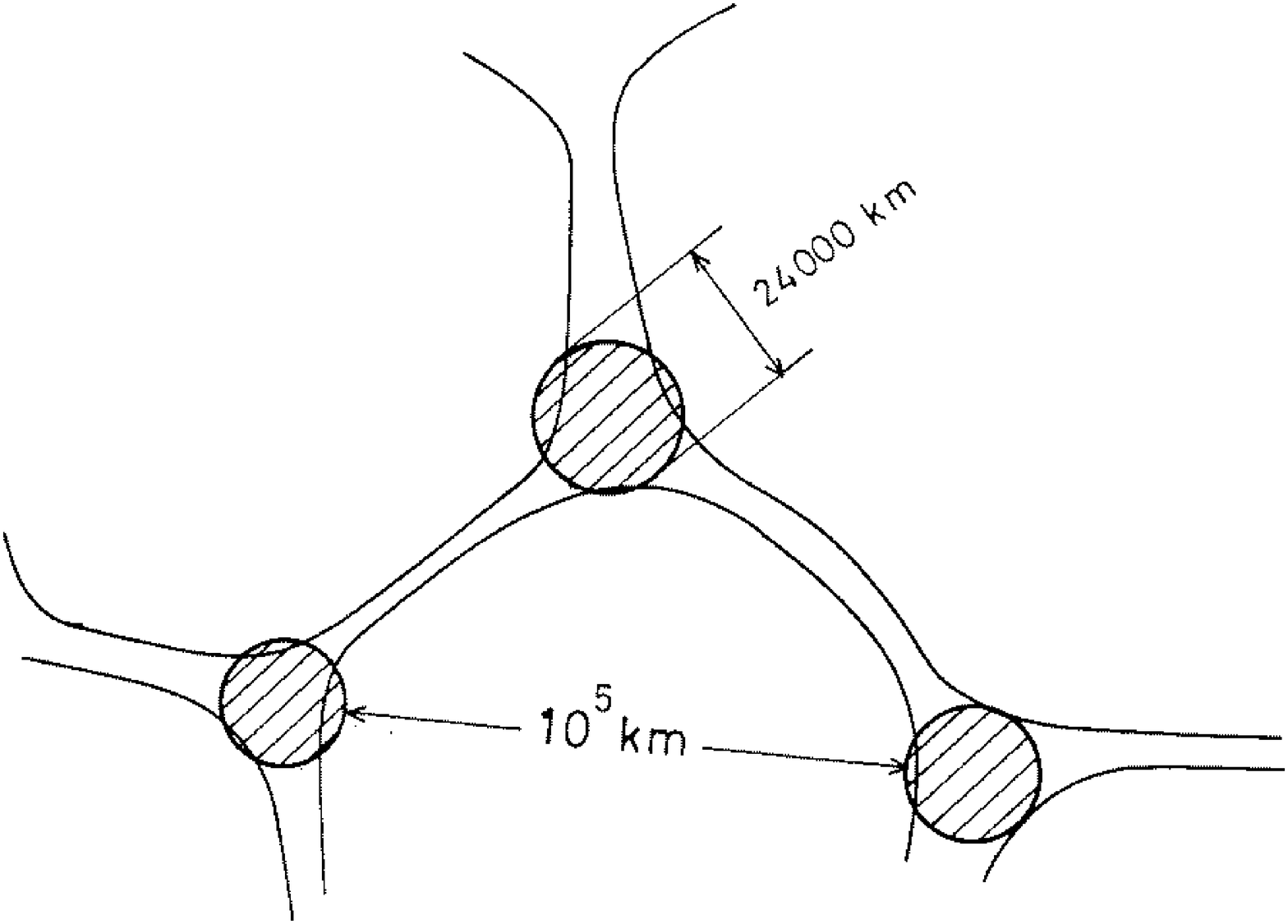}}
\caption{A sketch indicating positions of vertical magnetic flux tubes (shaded regions)
in a horizontal plane some distance above the tachocline, in the polar region
of the solar
convection zone.  The curved lines indicate boundaries of convection cells, where
the fluid is sinking.}
\end{figure}

Figure 2 gives a sketch of how the vertical flux tubes would appear distributed
in a horizontal section somewhat above the tachocline in the polar region.
The radii of the flux tubes are of order 12,000 km and their separations are
of order 85,000 km.  If the magnetic field inside flux tubes is of order 500~G,
then the mean magnetic is clearly of order 10~G.  The separations between the
flux tubes being quite large, at any time there would be only a few vertical
flux tubes in the polar region above the tachocline.  However, these few flux
tubes are expected to be quite long-lived, since the convective turnover time
in this region is of the order of months.  These flux tubes are expected to
be advected downward inside the tachocline, where they 
get stretched in the toroidal direction by the differential rotation.
What would the resulting distribution of toroidal flux
tubes look like? Since the vertical flux tubes above the tachocline had finite
lengths (as seen in Figure~1), we expect them to produce horizontal flux tubes
of finite length, with the magnetic field somewhat diffuse amongst these
concentrated flux tubes of finite length.  Figure~3 shows how the magnetic flux
in the tachocline may be distributed as seen from above.

\begin{figure}
\centerline{\includegraphics[height=7cm,width=7cm]{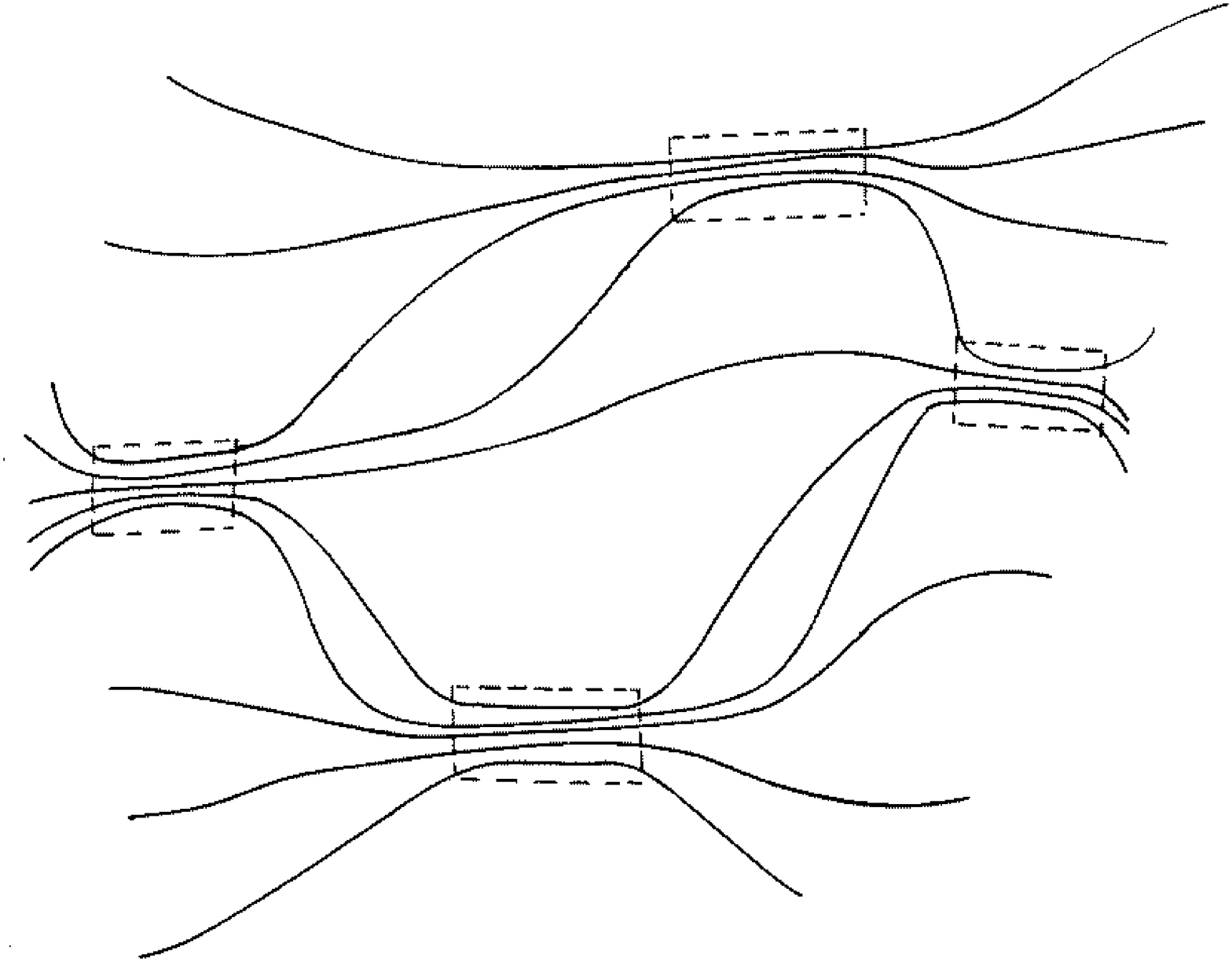}}
\caption{A sketch showing magnetic field lines at the bottom of the solar
convection zone, as seen from above.}
\end{figure}

\section{Equatorward advection of magnetic flux, magnetic buoyancy and 
formation of active regions}

We know that magnetic buoyancy is particularly destabilizing inside the convection
zone, but can be suppressed to a large extent in the sub-adiabatic region below
the bottom of the convection zone (see, for example, Parker, 1979, \S8.8).  The
usual scenario of active region formation is the following.  The toroidal flux
tubes remain stored in the stable layers below the bottom of the convection zone.
Occasionally a part of a toroidal flux tube may be pushed into the convection
zone, perhaps due to the disturbances caused by overshooting plumes getting into
the stable layers.  Once a part of a flux tube has entered the convection zone,
it becomes buoyant and rises in the form of a loop, eventually to produce
bipolar active regions (Moreno-Insertis, 1986; Choudhuri, 1989). In this view,
only a small portion of a flux tube makes it to the solar surface.  The major
part of the flux tube remains anchored inside the tachocline.  An active region
can thus be viewed to be like the tip of an iceberg, the much larger chunk of
the flux tube remaining embedded in the tachocline.  

If the recent model of Nandy and Choudhuri (2002) is correct, then we are led
to a radically different perspective.  According to their model, the toroidal flux tubes
are pushed into stable layers below the bottom of the convection zone immediately
after their creation within the tachocline at high latitudes.  However, when the
penetrating meridional circulation rises at the low latitudes and enters the
convection zone, it would carry all the flux tubes embedded in the fluid and
bring everything to the convection zone.  Thus the scenario of a flux tube sitting
within the tachocline and only a small part of it rising could not possibly be
correct if there is a meridional flow which penetrates below the tachocline and 
then sweeps through it at the low latitudes when it rises upward to enter the
convection zone. 

If the toroidal flux created within the tachocline is completely brought into
the convection zone at low latitudes, then how can we explain the formation of
active regions which seem to have roots anchored deep down?  We get a clue by
noting that magnetic field inside the tachocline is distributed as shown in 
Figure~3 and remembering that magnetic buoyancy is a function of magnetic field
strength.  Concentrated portions of a flux tube rise much more quickly than
other portions.  The acceleration due to magnetic buoyancy goes as $B^2$,
whereas the terminal velocity of rise (by the braking action of drag) goes
as $B$ (see, for example, Parker, 1979, \S8.7). When the magnetic configuration
shown in Figure~3 is brought into the convection zone by the meridional flow,
the concentrated portions of magnetic flux will rise rapidly in the form of
loops, whereas other diffuse portions will rise very slowly.  In this view,
the active regions appear anchored---not because other portions of the flux
tube down at the bottom of the convection zone remain anchored, but rather
because the flux tube ceases to exist as we follow magnetic field lines from
the active region to the bottom of the convection zone and the diffuse magnetic
field there rises much more slowly (mainly due to the advection 
by the upward moving meridional flow rather than magnetic buoyancy).
We can make a rough estimate of the diffuse field by assuming it to be
of the same order as mean $\Bp$.  Now, (5) should hold either for magnetic
fields in local regions or for mean magnetic fields, so we write
$$\ol{B}_{\phi} \approx 200 \, \ol{B}_r.$$
On taking $\ol{B}_r$ to be 10 G,
$$\ol{B}_{\phi} \approx 2000 \; \mbox{G}. \eqno(10)$$
If the diffuse field is of order 2000 G, then its rise time is of the order of
several years (Parker, 1975; Moreno-Insertis, 1983), whereas the rise time of
the concentrated portion with field 100,000 G is of the order of a few weeks.
The rise of the concentrated portions of magnetic field would be exactly like
the rise of the upper parts of flux loops in the earlier scenario of flux
tubes going into the stable layers where anchoring took place, and we believe 
that most of the important
results of numerical simulations based on the earlier scenario (Choudhuri, 1989;
D'Silva and Choudhuri, 1993; Fan, Fisher, and DeLuca, 1993; D'Silva and Howard 1993;
Caligari, Moreno-Insertis, and Sch\"ussler, 1995) should
carry over to the new scenario.

Although this new scenario may appear radical at the first sight, we are
not aware of any arguments against it.
The new scenario retains all the attractive aspects of the old scenario.
Additionally, the new scenario solves some problems which would find no easy
solution in the old scenario.  We now turn to these problems.

\subsection{The diffusion problem}

The magnetic field of the Sun reverses its direction every 11 years.  If
the active regions are merely tips of icebergs and major portions of flux
tubes are stored at the bottom of the convection zone, then these stored
flux tubes have to be destroyed in less than 11 years.  It is not clear
how this could have happened.  Certainly the resistivity of the plasma
is too small to achieve this.  The only possibility left is turbulent
diffusion.  If the magnetic field is as strong as 100,000 G, then even
turbulence will not be able to distort these flux tubes and mix them up
(Parker, 1993).
In kinematic models of the solar dynamo, one usually chooses some value
of turbulent diffusion which makes various things come out right.  But
there is no physical basis for assuming such turbulent diffusion if the
magnetic field is of order 100,000 G.  It has, therefore, remained a complete
mystery in the solar dynamo problem how the flux tubes with very strong
magnetic field at the bottom of the convection zone get destroyed in a few
years.

In the new scenario we are proposing, this problem is solved in one single
stroke.  We are suggesting that magnetic fields seen on the surface do not 
constitute the tip of an iceberg, but they are all there is to it.  There
is no further magnetic flux sitting at the bottom of the convection zone 
and waiting to disappear in a few years by some mysterious process.  The
concentrated parts of the magnetic field float to the surface due to magnetic
buoyancy, whereas the diffuse parts which rise slowly can be mixed up by
turbulence and destroyed.  According to the Babcock-Leighton idea, the
poloidal field at the surface is produced from the decay of active regions
(Babcock, 1961; Leighton, 1969), and we know that the poloidal field is then
freely advected poleward by the meridional circulation (Wang, Nash, and Sheeley, 1989;
Dikpati and Choudhuri, 1994, 1995).  If the active regions remained anchored to
flux tubes underneath the bottom of the convection zone, then it would have
been difficult for the poloidal field to be detached so easily and then to
be carried poleward.  In the scenario we are suggesting, it should be possible
for the poloidal field to be detached more easily.  Recent work by Longcope
and Choudhuri (2002) suggests that active regions do {\em not} get disconnected
at a shallow depth soon after the formation.  Presumably they eventually get
disconnected when the diffuse portions beyond the ends of the flux tubes
deep down in the convection zone get mixed up by turbulence.

\subsection{The magnetic tension problem}

In the model of Nandy and Choudhuri (2002), the toroidal flux tubes are carried
by the meridional flow from high latitudes to low latitudes through the stable
layers just below the bottom of the convection zone.  If the flux exists in
the form of a toroidal flux ring symmetric around the rotation axis, then we
know that it has to stretched if we want to take it from a high latitude to a low
latitude and such stretching involves working against magnetic tension.  For
a magnetic field of strength 100,000 G the magnetic tension would be very large,
even though the magnetic pressure would still be only about $10^{-5}$ of the
gas pressure and only a small perturbation in the gas pressure will be needed
to overcome the magnetic forces.  However, the kinetic energy density associated
with the meridional flow would be much less than the energy density of a
100,000 G magnetic field.  So,
one possible objection against the model of Nandy and Choudhuri (2002) is
that the meridional flow will not be able to carry the magnetic flux to low
latitudes by working against the large magnetic tension (Gilman, private 
communication).

This objection disappears in the scenario we are now proposing.  We are
suggesting that strong magnetic tubes are highly intermittent and are of finite
length beyond which the magnetic field becomes diffuse, i.e.\ there are no flux
rings going all the way round the rotation axis.  The magnetic tension goes as
$B^2$ and the tension of the diffuse field is much less. According to (10), the
diffuse field is of order 2000 G and the energy density associated with it is
about $10^5$ c.g.s.  Equating this to $\rho v^2/2$ and taking $\rho = 0.2$ gm
cm$^{-3}$, we get
$$v \approx 10 \; \mbox{m s}^{-1}. \eqno(11)$$
In other words, a meridional flow having a velocity larger than 10 m s$^{-1}$ at
the bottom of the convection zone would have more energy density than the diffuse
magnetic field and would be able to carry it.  The poleward flow velocity at the
surface is of order 20 m s$^{-1}$.  Although we have no direct knowledge of the
value of equatorward meridional flow at the bottom of the convection zone, it is
expected to be a few m s$^{-1}$ and most kinematic dynamo calculations assume such
values to get the best results.  Such a meridional flow may just have the right
strength to carry the diffuse magnetic field, without its tension being able to
pose a serious problem.  It would be difficult for the meridional flow to stretch
the bundles of concentrated field any further.  But the concentrated bundles of
flux can be carried to lower latitudes without further stretching if the diffuse
fields all around can get sufficiently stretched.  There is thus no need of doing
work against a very strong magnetic tension.

Since advection by meridional flow plays an important role in this model, perhaps
there may be a deep reason why the magnetic field may just be strong enough to
be carried by the meridional flow.  If the magnetic field were to become stronger,
then meridional flow might not have been able to advect the field, thereby reducing
the efficiency of the dynamo.  It is possible that this is the mechanism which
limits the growth of the magnetic field.  We plan to look at this provocative
question in near future.

\subsection{The Coriolis force problem}

We now turn to a related problem.  Apart from magnetic tension, there is another
force which resists the advection of flux tubes to low latitudes: the Coriolis
force.  Van Ballegooijen and Choudhuri (1988) studied the dynamics of an axisymmetric
flux ring at the bottom of the convection zone which is pushed towards lower
latitudes by an equatorward meridional flow.  As the flux ring is displaced
towards lower latitudes, it develops an internal azimuthal flow with respect to
the surrounding fluid in order to conserve angular momentum and the Coriolis
force arising out of that resists further displacement.  It was found by van
Ballegooijen and Choudhuri (1988) that various forces acting on the flux 
ring---magnetic buoyancy, magnetic tension, the Coriolis force and the drag
due to meridional flow---could keep the flux ring in stable equilibrium if certain
conditions were satisfied.  Only if there was efficient angular momentum exchange
between the flux ring and the surroundings, the Coriolis force would get reduced,
disrupting the stable equilibrium and allowing the flux ring to move further 
towards the lower latitudes.

With the magnetic configuration shown in Figure~3, we do not expect the Coriolis
force to pose a serious problem for the advection of magnetic flux.  Even if
azimuthal flows with respect to the surroundings are induced within the concentrated
portions of magnetic flux, such flows cannot continue beyond the concentrated flux
tubes and must stop where the flux tubes end.  Such flows, therefore, cannot become
very large.  In the regions of diffuse field, the magnetic field more or less would
fill up the whole volume.  Therefore, the question of relative velocity in regions
of magnetic flux concentration does not arise.  However, the meridional flow pushes
the whole fluid mass to lower latitudes and the fluid mass as a whole may start
rotating slower to conserve angular momentum, unless there is an efficient mechanism
of angular momentum removal.  One important question is whether the meridional
flow can penetrate below the tachocline into regions of solid-body rotation without
setting up a differential rotation there.  We still do not have a proper theoretical
model of the solar meridional circulation and the problem of angular momentum transport
associated with it is still very poorly understood (Durney, 2000).  If there is
meridional flow penetrating slightly below the tachocline, the advection of magnetic
field should not cause any additional problem if the magnetic configuration is
as shown in Figure~3.    
     
\section{Some characteristics of magnetic fields at the bottom of the convection
zone}

In the previous section, we have discussed some issues specifically related to the
recent model of Nandy and Choudhuri (2002).  We have seen how this model provides
an elegant answer to the question how the magnetic field gets destroyed in 11
years, which had otherwise remained completely mystifying. Now we again return 
to some generic considerations of the CDSD model which are not specific to the
model of Nandy and Choudhuri (2002).

We are concluding that the magnetic field in the stable layers just below the
convection zone exists in the form of intermittent flux concentrations having
internal field strength of about 100,000 G, with a much more diffuse field of
strength about 2000 G filling up the surrounding space.  Since convective turbulence
would be unable to twist flux tubes of strength 100,000 G and the traditional
$\alpha$-effect cannot be operative on these flux tubes, we had invoked the
Babcock-Leighton idea that the poloidal field is produced from the decay of active
regions at the surface.  Surface observations clearly show that this is happening
at the surface (Wang, Nash, and Sheeley, 1989).  However, if much of the volume at the bottom
of the convection zone is filled with magnetic field of 2000 G, then it is
certainly possible that the $\alpha$-effect works in regions other than
the interiors of flux tubes and the generation of the poloidal field from the diffuse
toroidal field of about 2000 G takes place at the bottom of the convection zone.
It may be noted that there had also been suggestions that various instabilities
associated with the strong field at the bottom of the convection zone may
drive the dynamo (Ferriz-Mas, Schmitt, and Sch\"ussler, 1994; Dikpati and Gilman, 2001).
These instability calculations, however, are extremely complicated and
their results are not always straightforward to interpret.  If much of the
volume is filled with diffuse magnetic field and the good old $\alpha$-effect
due to helical turbulence is present, then we can be much more sure that the
production of the poloidal field indeed takes place at the bottom of the
convection zone to some extent.

Choudhuri and Dikpati (1999) studied the evolution of the poloidal field
assuming that it has two sources: one at the bottom of the convection zone and
one at the surface where active regions decay. Certain aspects of the observational
data could be modeled particularly well by assuming two sources.  Perhaps
a complete model of the solar dynamo should also include both the sources
of the poloidal field, i.e.\ it should
incorporate features of both the CDSD model and the interface model
(Parker, 1993; Charbonneau and MacGregor, 1997).
Some worries have recently been expressed whether the CDSD model gives
the correct parity of the solar magnetic fields (Dikpati and Gilman, 2001;
Bonanno {\em et al.}, 2002).  It is not yet clear whether this problem is really
general and has to be taken sufficiently seriously (Charbonneau, private communication).
It is quite possible that the interface dynamo action on the diffuse field at
the bottom of the convection zone helps in fixing the polarity, whereas the
decay of tilted active regions produce the dominant poloidal field.  The 
model of Nandy and Choudhuri (2002) was based on calculations done in one
hemisphere.  We are now in the process of extending the model to the full
sphere with an additional layer of $\alpha$-effect at the bottom of the
convection zone (Nandy and Choudhuri, in preparation).

\def\Lp{L_{\phi}}

Let us now see whether we can draw some conclusions about the sizes and shapes of
the flux concentrations in the stable layers below the bottom of the convection
zone.  We are suggesting that the whole bundles of concentrated flux come up
in the form of active regions.  The typical length of a region of concentrated
magnetic field at the bottom of the convection zone should, therefore, be of the
same order as the sizes of active regions.  We take this length to be about
25,000 km.  We suggested in \S2--3 that vertical flux concentrations in the polar
region above the tachocline get stretched by the differential rotation to produce
the toroidal flux tubes.  Let $L_r$ be the typical length of vertical flux tube
in the polar region above the tachocline and let $\Lp$ be the length of the toroidal
flux tube.  Since the magnetic field runs along the axis of the flux tube, we would
expect in sufficiently simple situations that
$$\frac{\Lp}{L_r} = \frac{\Bp}{B_r}. $$
Taking $L_r \approx 50,000$ km (remember that the horizontal separations amongst
these flux tubes given by (9) is about 100,000 km) and making use of (5), we get
the enormous value 
$$\Lp \approx 10^7 \; \mbox{km}, \eqno(12) $$
which is close to 15 times solar radius!  This is certainly unrealistic.  Something
else must happen to fix the length $\Lp$ to much smaller values.  It is our
conjecture that probably the horizontal dimensions of convective plumes in the
overshoot layer under the bottom of the convection zone determine $\Lp$.  Once
a flux concentration forms with internal field strength 100,000 G, the convective
plumes would of course not be able to disturb it.  So, the convective plumes
must ensure that flux concentrations much larger their horizontal size do not
form at all.  This is what may happen if the meridional flow at the bottom of the
convection zone is not a spatially smooth flow, but is in the form of bursts
produced by the penetrating plumes.  The recent simulations of Miesch {\em et al.}
(2000) suggest that this may indeed be the case.
Still, we confess that we do not have a good understanding of what fixes
the observed length $\Lp$ to the rather low value of 25,000 km.  Not understanding
the reason behind this is perhaps the weakest link in the chain of arguments
we are giving.

We may assume that the cross-section of a typical flux tube in the polar region
above the tachocline would be circular, with the radius equal to about 12,000
km as given by (7).  When this flux tube is stretched in 
the azimuthal direction by differential rotation, the cross-section would tend
to become highly elliptical.  The semi-major axis in the $\theta$-direction
would still be 12,000 km, since there is no stretching in that direction.
In the $r$-direction, however, the flux tube would get compressed by the
factor 200 appearing in (5).  In other words, the semi-minor axis in the
$r$-direction would become only 60 km!  The cross-section of the flux tube
can retain the shape of a highly eccentric ellipse only if the flux tube has
no twist.  If there is twist, the magnetic tension associated with it would
lead to a circularization of the cross-section.  If the cross-section can
be circularized while the flux tube is still underneath the bottom of the convection
zone, it is easy to see that the radius would be about 1000 km.  We have no direct
way of knowing whether a flux tube at the bottom of the convection zone would
have sufficient twist to make its cross-section circular or whether it would
exist in the form of a highly flattened flux tube.  The only thing we can
say definitely is that when the flux tube emerges at the solar surface to
form active regions, no trace of the initial flattening is found any more.
If the flux tube cross-section is not circularized at the bottom of the convection
zone, then the circularization must take place during the rise phase.  There
are ways a flux tube can acquire twist during its rise, leading to the
circularization of cross-section.  The twist of solar flux tubes is a very
important subject and we discuss it in the next section.

\section{Twist and helicity}

\subsection{Observed current helicity of active regions and a theoretical
explanation}

The magnetic loops above active regions appear twisted and several groups
independently have established that active regions in the northern hemisphere
of the Sun predominantly have negative current helicity $\jb.\Bb$ 
(Seehafer, 1990; Pevtsov, Canfield, and Metcalf, 1995; 
Abramenko, Wang, and Yurchishin, 1997; Bao and Zhang, 1998; 
Pevtsov, Canfield, and Latushko, 2001).  There are ways in which a
flux tube can pick up some twist while rising through the convection zone,
through interactions with helical turbulence in the surrounding region
(Longcope, Fisher, and Pevtsov, 1998).  One very important question 
is whether this twist
could be the result of dynamo action.  We present below a very simple
and elegant argument
that Babcock-Leighton type dynamos indeed are expected to produce negative
current helicity in the northern hemisphere.  Since this argument is so
disarmingly simple, we have been wondering if this argument occurred to
somebody before us.  We have, however, not seen this particular argument
presented anywhere.  So, with some hesitation, we present this below as an
original argument.

\def\vb{{\bf v}}

Figure~4 shows a section of the convection zone in the northern hemisphere with
the $\phi$-direction into the paper.  The equator is towards the right side 
and the pole towards the left side.  Suppose we look at the system at a time
when the concentrated flux tubes at the bottom of the convection zone have
positive $\Bp$, i.e.\ the magnetic fields in the flux tubes are going into
the paper.  Some of these flux tubes would rise to the surface and produce the
poloidal field at the surface by the Babcock-Leighton process.  Keeping in
mind that the leading sunspot in an active region is found nearer the equator,
it is not difficult to see that flux tubes with positive $\Bp$ would give rise
to a poloidal field with field lines going in the clockwise direction.
This also follows from mathematical considerations.  The standard equation
describing the evolution of the poloidal field is
$$\frac{\pa A}{\pa t} + \frac{1}{s}(\vb.\nabla)(s A)
= \eta \left( \nabla^2 - \frac{1}{s^2} \right) A + \alpha \Bp. \eqno(13)$$
It is well known that the essence of the Babcock-Leighton process is
captured by taking $\alpha$ concentrated near the solar surface with a
positive value in the northern hemisphere (see, for example, Nandy and
Choudhuri, 2001).  If $\alpha$ near the surface is positive and $\Bp$ inside
the flux tubes coming to the surface by magnetic buoyancy is also positive,
then clearly (13) suggests the production of positive $A$ at the surface.
The poloidal field is given by $\Bb_p = \nabla \times (A {\bf e}_{\phi})$
and it is straightforward to see that the field lines would be clockwise
around a region of positive $A$.

\begin{figure}
\centerline{\includegraphics[height=7cm,width=7cm]{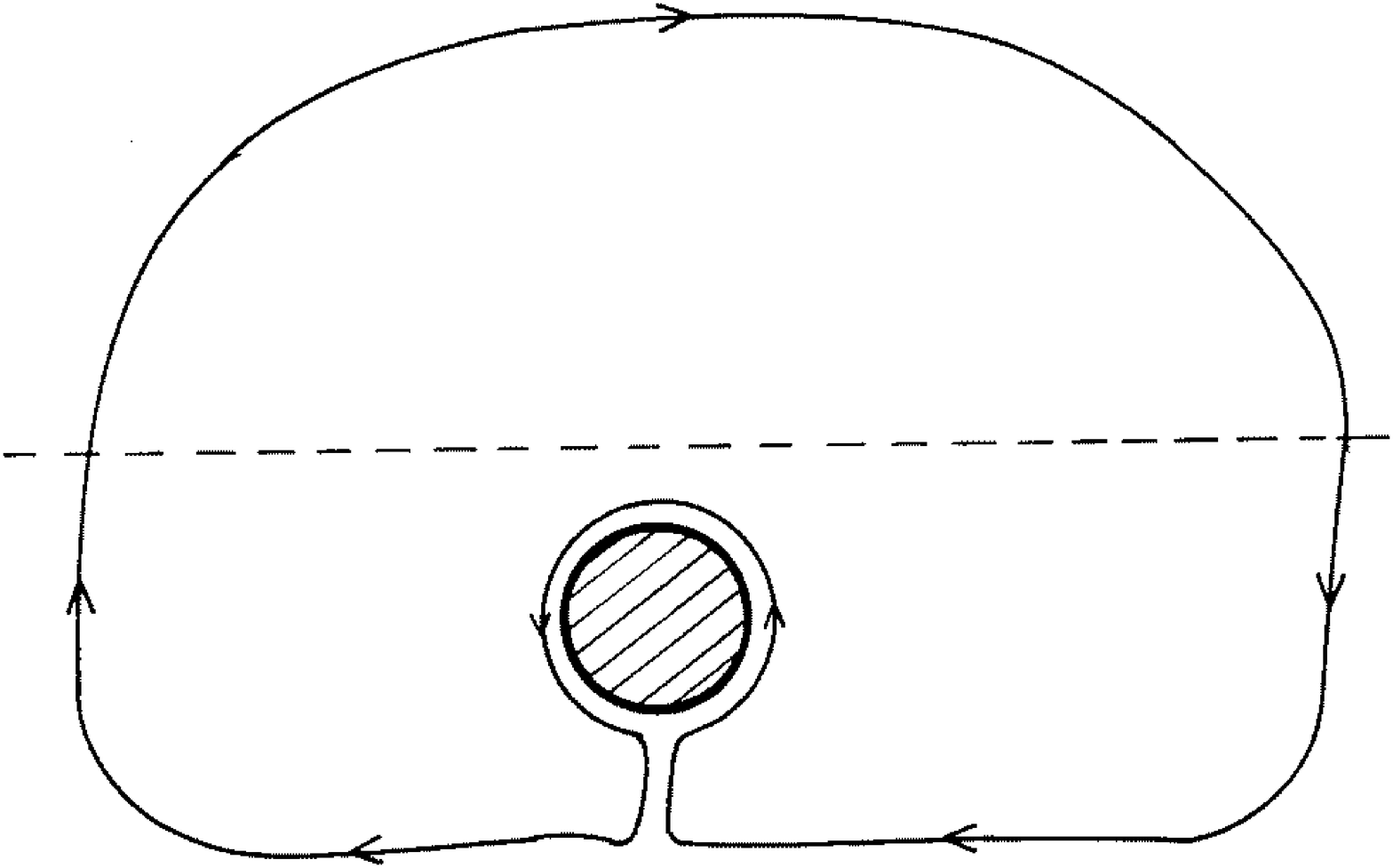}}
\caption{Magnetic field lines around a rising flux tube, with the dashed
line indicating the solar surface.  The flux tube, shown by the shaded
circle, has magnetic field going into the paper, and it is rising in a
region of clockwise poloidal field.}
\end{figure}
Now suppose a new flux tube rises to the surface and moves into the region
of clockwise poloidal field.  Since we are dealing with a high magnetic
Reynolds number situation, the rising flux tube would not be able to cut the
poloidal field lines easily and we expect that the poloidal field lines would
be pushed  by the rising flux tube as shown in Figure~4.  Eventually the field
lines left in the wake behind the flux tube would reconnect and we shall be
left with an anti-clockwise twist around the flux tube.  With $\Bp$ inside the
flux tube directed into the paper, it is easy to see that an anti-clockwise twist
implies {\em negative} current helicity.  If we had carried out our arguments
with flux tubes having $\Bp$ out of the paper (i.e.\ negative $\Bp$), then
the poloidal field lines would have been anti-clockwise and the twist around a
rising flux tube clockwise, again leading to negative current helicity.
When a flux tube in the northern hemisphere rises to the surface in a region
of pre-existing poloidal field {\it created by flux tubes of similar type}
which erupted earlier, we conclude that the flux tube acquires negative 
current helicity.  We thus find that the observed 
predominant negative current helicity
of active regions in the northern hemisphere has a very simple explanation
within the framework of the Babcock-Leighton dynamo model. It is easy to
check that our argument applied to the southern hemisphere would imply
positive current helicity, in accordance with observations. 

Although the current helicity of active regions in the northern hemisphere
is predominantly negative, many active regions are found with positive
current helicity also (Pevtsov, Canfield, and Metcalf, 1995; 
Abramenko, Wang, and Yurchishin, 1997; Bao and Zhang, 1998).
An extremely important question is whether this is merely statistical
fluctuation or whether there is some systematic aspect in it, i.e.\ if
the positive current helicity is preferentially found in certain latitudes
at certain times.  More extensive data analysis will be required to
settle this question.  On theoretical grounds, we expect that the flux
tubes may have positive current helicity systematically in certain 
latitudes at certain times.  The argument we had given above hinges
crucially on the fact that the emerging flux tube has to come up in
a region where the poloidal field has been earlier created by flux tubes of the
same kind, i.e.\ a flux tube with positive $\Bp$ has to come up in a
region of positive $A$.  We know that flux tubes at the bottom of the
convection zone are advected equatorward by the equatorward meridional
flow there, whereas the poloidal field at the surface is advected poleward
by the poleward meridional flow there.  So, during certain phases of the
solar cycle, there may be a situation in certain latitudes that $A$ at
the surface has a sign which is opposite to the sign of $\Bp$
of the flux tubes at the bottom of the convection zone.  In such a
situation, the active regions would be created with positive current
helicity.  A systematic study of the observed signs of current helicity
in different latitudes in different phases of the solar cycle should
throw important light on the nature of the solar dynamo.

\subsection{Helicities at small and large scales}

Now that we have seen how a rising flux tube in the northern hemisphere
would pick up negative helicity, let us consider how that would appear
from a mean field point of view.  In a mean field approach, we average
over small length scales such as the cross-section of the flux tubes.
On taking average over scales larger than the flux tube cross-section,
the anti-clockwise twist just outside the flux tube 
seen in Figure~4 would contribute zero.  So we would
conclude the large scale twist to be clockwise, which combined 
with positive $\Bp$ would give us positive current helicity in the
northern hemisphere.  We thus find that the helicity at the large scale
is opposite in sign to the helicity of flux tubes (which we consider the
small scale). 

It is well known that the dynamo process generates helicity.  While discussing
the helicity production in the dynamo process, it is more useful to think
in terms of magnetic helicity $\Ab.\Bb$ rather than current helicity $\jb.\Bb$.
It is true that the definition of $\Ab$ is not unique due to gauge freedom
and consequently there is an uncertainty in defining magnetic helicity in
a local region.  However, since the volume 
integral of magnetic helicity is connected
with the topology of magnetic field (see, for example, Choudhuri, 1998),
it is still an extremely useful concept, and once the gauge is fixed, we
can define magnetic helicity. Unless a magnetic field has a very
pathological configuration, we can normally assume that the current
helicity and the magnetic helicity would have the same sign. 
It is easy to show that a dynamo with positive $\alpha$-effect produces
positive magnetic helicity of the mean fields (Seehafer, 1996).   For
the benefit of the readers not very familiar with the subject, 
we give a simple proof in the Appendix that, in the absence of
dissipation, a dynamo in a finite volume with closed field lines and with
a constant $\alpha$ would generate magnetic helicity $H= \int \Ab.\Bb dV$
at the rate
$$\frac{d H}{d t} = 2 \alpha \int B^2 d V, \eqno(14)$$
clearly showing that a positive $\alpha$ implies the production of positive
$H$ within the mean field framework.

Since magnetic helicity is connected with the topology of field lines, 
it is not possible to change the value of magnetic helicity in time
scales shorter than the resistive decay time (which is very large
for astronomical systems).
Then how does the dynamo produce magnetic helicity of the mean field?  It
has been shown by Seehafer (1996) that a dynamo generates magnetic helicity
at the large scales by transferring magnetic helicity of equal and opposite
amount to the small scales, thus ensuring that the total amount of magnetic
helicity does not change.
In Figure~4 we have presented a clear physical picture
of how this happens in the Babcock-Leighton framework.  A positive $\alpha$
in the northern hemisphere of the Sun would produce positive helicity at
the large scales and hence negative helicity at the small scales.  By
identifying the helicity of the flux tubes as the helicity associated with
small scales, we get an understanding of why the active regions in the
northern  region should have predominantly negative helicity, in accordance
with the observations.

\subsection{The ultimate fate of helicity}

A dynamo with positive $\alpha$ keeps on piling negative helicity at the small
scales.  If the dynamo cannot get rid of this negative helicity at the small
scales, it eventually gets quenched.  This is believed to be the reason for
dynamo quenching found in some numerical simulations (Blackman and Field,
2000). The helicity constraint is very severe in $\alpha^2$ dynamos.  Although
it is less severe for $\alpha \Omega$ dynamos, it still has to be reckoned
with (Brandenburg, Bigazzi, and Subramanian, 2001).  In the CDSD models, the
toroidal and the poloidal fields generated in spatially separated regions,
and are not inter-linked when they are generated.  The helicity is produced
when toroidal flux tubes generated at the bottom of the convection zone move
into regions of poloidal field, as seen in Figure~4. Still, the Sun has to get
rid of the helicity continuously, to ensure that
the helicity constraint does not cause any problem for the solar dynamo.  

Once a flux tube emerges through the solar surface, the upper portion of it
becomes a coronal loop with a cross-section much larger than that below the
surface.  Parker (1974) studied the magnetohydrostatic equilibrium of flux
tubes with variable cross-section and showed that most of the twist would
get concentrated in the region where the cross-section is maximum.  Even if a
solar flux tube initially had twist underneath the solar surface, after emergence
through the surface the twist would presumably propagate in the form of 
torsional Alfven waves to be eventually concentrated in the uppermost part
of the loop.  Thus the negative helicity dumped by the solar dynamo 
at the small scales in the northern hemisphere ultimately makes its way
to the upper parts of coronal loops.  Magnetic filaments in the corona
are indeed known to have opposite twists in the two hemispheres 
(Rust and Kumar, 1994).  Once the twisted part of the flux
tube rises through the solar surface and goes out of the convection zone,
it is clear that there is no helicity left within the convection zone---either
positive helicity at the large scales or negative helicity at the small
scales.  So there would be no problem in the convection zone arising out
of the helicity constraint.

The Sun eventually has to get rid of the helicity at the tops of coronal
loops by various coronal processes.  If the loop becomes sufficiently
twisted, it may eventually become unstable and lead to such coronal phenomena
as flares or CMEs.  It has been conjectured by Low (1994) that
CMEs may constitute a mechanism by which the Sun gets rid of its excess 
helicity.  It is, therefore, conceivable that the helicity of the Sun
is ultimately carried away by the solar wind.
When we consider the connection between the solar dynamo and the flux tubes,
many different pieces of the jigsaw puzzle seem to fall in place.

\section{Conclusion}

To the best of our knowledge, no serious attempt has been made previously
to reconcile the existence of flux tubes with the results of mean field
dynamo theory.  We suggest a possible procedure here: first solve the mean
field dynamo equation to figure out how the mean field behaves
and then use other considerations (such as results
of flux tube calculations) to figure out where flux tubes would exist and
what would be their nature.  Only the second step is attempted in this
paper, based on the solutions of CDSD model presented by various previous
authors.  This paper is quite unlike the other recent papers of the
present author where usually detailed calculations are presented.  This paper
is based more on order-of-magnitude estimates and general arguments.  The 
nature of the subject is such that it is important first to evolve a plausible
scenario based on such estimates and arguments before any detailed calculation
is attempted. We have, in fact, indicated a few possibilities of detailed
calculations to make our picture more complete.  We hope that such calculations
will be done in near future.

The polar field of the Sun appears reasonably diffuse.  It is not clear to
what extent this field is concentrated in fibril flux tubes, but it is
certainly not organized in flux concentrations as large as sunspots.
In the CDSD model, this polar field is advected by the meridional circulation
to the tachocline, where it is stretched to produce the strong toroidal
field.  Our surface observations are the following: the polar field at
the time of subduction below the surface is reasonably smooth and the
active regions at the time of emergence appear in the form of flux tubes.
Hence the magnetic field must get organized into flux concentrations 
at some stage during the subsurface processes of advection of the polar 
field to the tachocline, the stretching by differential rotation and the
subsequent rise of the toroidal field by magnetic buoyancy.  Simple
and straightforward considerations based on the dynamo equation conclusively
rule out the possibility that the magnetic field could be smooth at the
bottom of the convection zone, flux tubes breaking away from this smooth
field due to some instability and rising thereafter.
The polar is field is about 10 G, whereas
flux tube calculations suggest that the sunspot-forming toroidal field
at the bottom of the convection zone must be of order 100,000 G.  An
order-of-magnitude estimate shows that the differential rotation in the
tachocline is not sufficient to stretch a 10 G poloidal field into a
uniform 100,000 G toroidal field.  From such estimates, we have been able to
draw some conclusions about the volume filling factors of flux tubes.

We have suggested that the organization into flux tubes takes place while
the polar field is being advected downward into the tachocline.  We have
estimated these flux tubes in the polar region above the tachocline to
have radii of order 12,000 km and to have mean separations of order 85,000 km.
Certainly not too many such flux tubes can exist at a certain time.  But
we expect them to be long-lived and presumably some of these flux concentrations
eventually become active regions after a few years, after being stretched in
the toroidal direction in the tachocline while they are advected to lower
latitudes.  Our estimate of filling factor should be of interest to
helioseismologists who are looking for signatures of magnetic field at
the bottom of the convection zone.  In a decade or two, perhaps the techniques
of helioseismology may become sophisticated enough to study the formation of
flux tubes above the tachocline and then follow them as they get titled
in the toroidal direction and are advected to lower latitudes, eventually
to emerge as active regions.  The model of Nandy and Choudhuri (2002) imply
that the active regions could not be like tips of icebergs, constituting
a small part of the toroidal field sitting at the bottom of the convection
zone.  Rather, all the toroidal flux at the bottom of the convection zone
has to come up when the penetrating meridional flow rises.  This solves the
mystery of how the strong 100,000 G field gets annihilated in 11 years.
We also point out that the intermittent nature of the magnetic field in the
tachocline makes it possible for the meridional flow to advect it to the lower
latitudes, without being resisted by the magnetic tension or the Coriolis
force.

The major uncertainty in our scenario is that we do not understand what 
determines the lengths of toroidal flux concentrations at the bottom of
the convection zone.  We have suggested that the horizontal widths of
plumes penetrating below the bottom may have something to do with this.
Perhaps a numerical simulation to study how the magnetic configuration shown
in Figure~2 would get advected into the tachocline by intermittent plumes
would throw more light on this question.

During the last decade, considerable work has been done on the current helicities
of active region.  Our model gives a disarmingly simple explanation of the
observed current helicity.
Flux tubes from the bottom of the convection zone rise to the region where there
exists a poloidal field created by similar flux tubes earlier and they get a twist
in this process.  This explanation matches the sign of the observed current
helicity.  A fundamental question in dynamo theory is the relation between
helicities at large and small scales, as well as the ultimate fate of the helicity.
When we combine flux tube considerations with considerations of dynamo theory,
we seem to get very natural answers to some of these intriguing questions. 

\section*{Acknowledgements}

This work would not have been possible without my ongoing collaboration with
Dibyendu Nandy, with whom I carried out the dynamo calculations and discussed
many of the issues presented in this paper.  I also wish to thank Axel Brandenburg,
Dick Canfield, Paul Charbonneau, Bernard Durney, Peter Gilman, Dana Longcope,
Gene Parker, G\"unther R\"udiger, Manfred Sch\"ussler and Kandaswamy 
Subramanian. Many of the ideas presented in this paper started taking shape
in my mind during my many discussions with them over the last few years.
I am grateful to the Alexander von Humboldt Foundation for supporting a
visit to the Max-Planck-Institut f\"ur Aeronomie in Lindau during the summer
of 2002, when this work was seriously initiated.


\section*{Appendix. Magnetic helicity generation in dynamo process}

We consider dynamo action in a volume bounded by a surface through which
fluid elements or magnetic field lines do not pass (i.e.\ 
both $\vb$ and $\Bb$ on the boundary are zero).  Let the $\alpha$ coefficient
be constant within this volume and let the diffusion coefficient be zero.
Then the dynamo equation is given by
$$ \frac{\pa \Bb}{\pa t} = \nabla \times (\vb \times \Bb) + \nabla \times (\alpha
\Bb).  \eqno(15)$$
We choose the gauge of the vector potential such that it satisfies the equation
$$ \frac{\pa \Ab}{\pa t} = \vb \times \Bb + \alpha\Bb. \eqno(16)$$
The rate of change of the magnetic helicity $H = \int \Ab. \Bb dV$ is given
by
$$ \frac{dH}{dt} = \int \frac{\pa \Ab}{\pa t}. \Bb \, dV + 
\int \Ab. \frac{\pa \Bb}{\pa t} \, dV.$$
On substituting from (15) and (16), we get
$$ \frac{dH}{dt} = \alpha \int B^2 \, dV + 
\int \Ab. [\nabla \times (\vb \times \Bb + \alpha\Bb)] \, dV. \eqno(17) $$
The second integral on the right hand side can be written in the following way
$$ \int(\vb \times \Bb + \alpha\Bb). \nabla \times \Ab \, dV - 
\int \nabla. [\Ab \times (\vb \times \Bb + \alpha\Bb)] \, dV. \eqno(18)$$
The second term here is a volume integral of a divergence and can be
converted into a surface integral by Gauss's theorem.  It is easy to
see that this term would vanish if $\vb$ and
$\Bb$ are zero on the surface.  Remembering that $\nabla \times \Ab = \Bb$,
we note that the first term in (18) gives $\alpha \int B^2 d V$.
It then follows from (17) that
$$ \frac{dH}{dt} = 2 \alpha \int B^2 \, dV. $$
We thus see that dynamo action generates magnetic helicity having the
same sign as $\alpha$.

\section*{References}

\def\apj{{\it Astrophys.\ J.}}
\def\mnras{{\it Mon.\ Notic.\ Roy.\ Astron.\ Soc.}}
\def\sol{{\it Solar Phys.}}
\def\aa{{\it Astron.\ Astrophys.}}
\def\gafd{{\it Geophys.\ Astrophys.\ Fluid Dyn.}}
\def\bi{}

Abramenko, V.\ I., Wang, T., and Yurchishin, V.B.: 1997, \sol\ {\bf 174}, 291.

{
{
\leftskip=20pt
\parindent=-\leftskip

  Babcock, H.\ W.: 1961, \apj\ {\bf 133}, 572.

\bi
   Bao, S.\ and Zhang, H.: 1998, \apj\ {\bf 496}, L43.

 \bi
  Blackman, E.\ G.\ and Field, G.\ B.: 2000, \apj\ {\bf 534}, 984.

 \bi
  Bonanno, A., Elstner, D., R\"udiger, G., and Belvedere, G.: 2002, \aa\ 
   {\bf 390}, 673.

 \bi
  Brandenburg, A., Bigazzi, A., and Subramanian, K.: 2001, \mnras\ {\bf 325}, 685.

 \bi
  Braun, D.\ L.\ and Fan, Y.: 1998, \apj\ {\bf 508}, L105.

  \bi
  Caligari, P., Moreno-Insertis, F., and Sch\"ussler, M.: 1995, \apj\
  {\bf 441}, 886.

   \bi
  Charbonneau, P.\ and MacGregor, K.\ B.: 1997, \apj\ {\bf 486}, 502.

 \bi
  Choudhuri, A.\ R.: 1989, \sol\ {\bf 123}, 217.

  \bi
  Choudhuri, A.\ R.: 1998, {\it The Physics of Fluids and Plasmas:
  An Introduction for Astrophysicists}, Cambridge University Press.
 
 \bi
  Choudhuri, A.\ R.: 2002, in B.\ N.\ Dwivedi (ed.), {\it The
  Dynamic Sun}, Cambridge University Press, p.\ 103. 

 \bi
  Choudhuri, A.\ R.\ and Dikpati, M.: 1999, \sol\ {\bf 184}, 61.

 \bi
  Choudhuri, A.\ R.\ and Gilman, P.\ A.: 1987, \apj\ {\bf 316}, 788.

 \bi
  Choudhuri, A.\ R., Sch\"ussler, M., and Dikpati, M.: 1995, \aa\
  {\bf 303}, L29.

 \bi
  Dikpati, M.\ and Charbonneau, P.: 1999, \apj\ {\bf 518}, 508.

 \bi
  Dikpati, M.\ and Choudhuri, A.\ R.: 1994, \aa\ {\bf 291}, 975.

 \bi
  Dikpati, M. and Choudhuri, A.\ R.: 1995, \sol\ {\bf 161}, 9.

 \bi
 Dikpati, M. and Gilman, P.\ A.: 2001, \apj\ {\bf 559}, 428.

 \bi
  D'Silva, S.\ and Choudhuri, A.\ R.: 1993, \aa\ {\bf 272}, 621.

 \bi
  D'Silva, S.\ and Howard, R.\ F.: 1993, \sol\ {\bf 148}, 1.

 \bi
  Durney, B.\ R.: 1995, \sol\ {\bf 160}, 213.

 \bi
  Durney, B.\ R.: 1996, \sol\ {\bf 166}, 231.

 \bi
  Durney, B.\ R.: 1997, \apj\ {\bf 486}, 1065.

 \bi
  Durney, B.\ R.: 2000, \apj\ {\bf 528}, 486.

 \bi
  Fan, Y., Fisher, G.\ H., and DeLuca, E.\ E.: 1993, \apj\ {\bf 405}, 390.

 \bi
  Ferriz-Mas, A., Schmitt, D., and Sch\"ussler, M.: 1994, \aa\
  {\bf 289}, 949.

 \bi 
  Giles, P.\ M., Duvall, T.\ L., and Scherrer, P.\ H.: 1997, {\it Nature} {\bf 390}, 52.

 \bi
  Gilman, P.\ A.: 1983, {\it Astrophys.\ J.\ Suppl.} {\bf 53}, 243.

 \bi
  Glatzmaier, G.\ A.: 1985, \apj\ {\bf 291}, 300.

 \bi 
  K\"uker, M., R\"udiger, G., and Schultz, M.: 2001, \aa\ {\bf 374}, 301.

 \bi
  Leighton, R.\ B.: 1969, \apj\ {\bf 156}, 1.

 \bi
  Longcope, D.\ and Choudhuri, A.\ R.: 2001, \sol\ {\bf 205}, 63.

 \bi
  Longcope, D.\ and Fisher, G.\ H.: 1996, \apj\ {\bf 458}, 380.

 \bi
  Longcope, D., Fisher, G.\ H., and Pevtsov, A.\ A.: 1998, \apj\ {\bf 507}, 417.

 \bi
  Low, B.\ C.: 1994, {\em Phys.\ Plasmas} {\bf 1}, 1684.

 \bi 
  Miesch, M.\ S., Elliott, J.\ R., Toomre, J., Clune, T.\ L., Glatzmaier, G.\ A., and
  Gilman, P.\ A.: 2000, \apj\ {\bf 532}, 593.

  \bi
  Moffatt, H.\ K.: 1978, {\it Magnetic Field Generation in Electrically
  Conducting Fluids}, Cambridge University Press.

 \bi
  Moreno-Insertis, F.: 1983, \aa\ {\bf 122}, 241.

 \bi
  Moreno-Insertis, F.: 1986, \aa\ {\bf 166}, 291.

 \bi
  Nandy, D.\ and Choudhuri, A.\ R.: 2001, \apj\ {\bf 551}, 576.

\bi 
  Nandy, D.\ and Choudhuri, A.\ R.: 2002, {\em Science} {\bf 296}, 1671.

 \bi
  Parker, E.\ N.: 1955, \apj\ {\bf 122}, 293.

 \bi
  Parker, E.\ N.: 1974, \apj\ {\bf 191}, 245.

 \bi
  Parker, E.\ N.: 1975, \apj\ {\bf 198}, 205.

 \bi
  Parker, E.\ N.: 1979, {\it Cosmical Magnetic Fields}, Oxford
  University Press.

 \bi
  Parker, E.\ N.: 1993, \apj\ {\bf 408}, 707.

 \bi
  Pevtsov, A.\ A., Canfield, R.\ C., and Latushko, S.\ M.: 2001, \apj\ 
  {\bf 549}, L261.
 
 \bi
  Pevtsov, A.\ A., Canfield, R.\ C., and Metcalf, T.\ R.: 1995, \apj\ 
  {\bf 440}, L109.

 \bi
  Rust, D.\ M.\ and Kumar, A.: 1994, \sol\ {\bf 155}, 69.

 \bi
  Seehafer, N.: 1990, \sol\ {\bf 125}, 219.

 \bi
  Seehafer, N.: 1996, {\em Phys.\ Rev.\ E} {\bf 53}, 1283.

  \bi
  Steenbeck, M., Krause, F., and R\"adler, K.-H.: 1966, {\it Z.\ 
  Naturforsch.} {\bf 21a}, 1285.

 \bi
  van Ballegooijen, A.\ A.\ and Choudhuri, A.\ R.: 1988, \apj\
  {\bf 333}, 965.

\bi
  Wang, Y.-M., Nash, A.\ G., and Sheeley, N.\ R.: 1989, \apj\
  {\bf 347}, 529.

}
}
\end{document}